\documentclass[a4paper,aps,prd,10pt,preprintnumbers,showpacs,twocolumn,superscriptaddress,nofootinbib,amsmath,amssymb]{revtex4-1}
\usepackage{cases}
\usepackage{graphicx}
\usepackage[utf8]{inputenc}
\usepackage[T1]{fontenc}
\usepackage{cmap}
\usepackage[normalem]{ulem} 
\def\imo{i}
\def\re#1{Re(#1)}

\def\K{{\cal K}}

\begin{document}
\title{Quasinormal modes of black holes in 5D Gauss-Bonnet gravity combined with non-linear electrodynamics}
\author{M. S. Churilova}\email{wwrttye@gmail.com}
\affiliation{Institute of Physics and Research Centre of Theoretical Physics and Astrophysics, Faculty of Philosophy and Science, Silesian University in Opava, CZ-746 01 Opava, Czech Republic}
\author{Z. Stuchlík}\email{zdenek.stuchlik@fpf.slu.cz}
\affiliation{Institute of Physics and Research Centre of Theoretical Physics and Astrophysics, Faculty of Philosophy and Science, Silesian University in Opava, CZ-746 01 Opava, Czech Republic}
\begin{abstract}
Quasinormal modes of black holes were previously calculated in a non-linear electrodynamics and in the Gauss-Bonnet gravity theory. Here we take into consideration both of the above factors and find quasinormal modes of a (massive) scalar field in the background of a black hole in the five-dimensional Einstein-Gauss-Bonnet gravity coupled to a non-linear electrodynamics having Maxwellian weak-field limit. For the non-linear electrodynamics we considered the high frequency (eikonal) regime of oscillations analytically, while for the lower multipoles the higher order WKB analysis with the help of Pad\'{e} approximants and the time domain integration were used. We found that perturbations of a test scalar field violate the inequality between the damping rate of the least damped mode and the Hawking temperature, known as the Hod's proposal. This does not exclude the situation in which gravitational spectrum may restore the Hod's inequality, so that  only the analysis of the full spectrum, including gravitational perturbations, will show if the quasinormal modes we found here for the scalar field can be a counterexample to the Hod's conjecture or not. We also revealed that in such a system, which includes the higher curvature corrections and non-linear electrodynamics, for perturbations of a massive scalar field there exists the phenomenon of the arbitrary long lived quasinormal modes - quasiresonances.
\end{abstract}
\maketitle

\section{Introduction}

Perturbations and proper oscillation frequencies (quasinormal modes) of black holes have been thoroughly studied during past years \cite{Kokkotas:1999bd}, \cite{Berti:2009kk}, \cite{Konoplya:2011qq}. Although the recent observation of the gravitational waves \cite{Abbott:2016blz} is consistent with the Einstein gravity  \cite{TheLIGOScientific:2016src}, it leaves large uncertainties in determination of the geometries of observed black holes, which gives the room to various alternative theories \cite{Konoplya:2016pmh}. One of such theories is predicted by the low-energy limit of heterotic string theory and implies higher curvature corrections to the Einstein gravity, given in the form of the second order curvature (Gauss-Bonnet) term. Various observable effects in the background of astrophysically viable models of black holes with Gauss-Bonnet corrections have been recently considered \cite{Nampalliwar:2018iru,Pani:2009wy,Witek:2018dmd}.

Non-linear electrodynamics was introduced to avoid the divergence of electron's self-energy in Maxwell electrodynamics. In the theories coupled to a non-linear electrodynamics the Lagrangian density is defined in such a way that the electric field strength has an upper bound and the self-energy of the point-like charges is finite. Various optical phenomena in the background of black holes in the presence of non-linear electrodynamics were considered in \cite{SS1,SS2} and some constraints were suggested. Although the electric charge of galactic black holes is not believed to be large, it is an important parameter when considering mini-black holes and higher dimensional scenarios \cite{Zajacek:2018vsj}.

Quasinormal modes have been calculated either in various theories of non-linear electrodynamics  \cite{Breton:2017hwe,Fernando:2012yw,Fernando:2005gh,Li:2014fka,Li:2014fka,Becar:2015kpa,Breton:2016mqh,Toshmatov:2015wga,Toshmatov:2018ell,Toshmatov:2018tyo} or in Gauss-Bonnet theories in four and higher dimensions in various contexts, including astrophysics, higher dimensional gravity and AdS/CFT correspondence \cite{Konoplya:2004xx,Abdalla:2005hu,Konoplya:2008ix,Cuyubamba:2016cug,Konoplya:2017ymp,Gonzalez:2017gwa,Gonzalez:2017gwa,
Konoplya:2019hml,Zinhailo:2019rwd,Pani:2009wy,Konoplya:2017zwo}. Perturbations of black holes in the Gauss-Bonnet theory has a number of distinctive features. The decayed quasinormal modes indicate the stability of black hole, which is changed to the instability when the Gauss-Bonnet coupling constant is not small \cite{Konoplya:2008ix,Cuyubamba:2016cug,Andrade:2016rln}, and the instability develops at high multipole numbers $\ell$, leading to the breakdown of the well-posedness of the initial value problem. The mode which induces the instability is purely imaginary and non-perturbative in the Gauss-Bonnet coupling constant \cite{Gonzalez:2017gwa} - \cite{Grozdanov:2016vgg}  which, in some cases, is important for the  description of quantum systems at intermediate coupling via gauge/gravity duality \cite{Grozdanov:2016vgg}. For asymptotically flat black holes in the Einstein-Gauss-Bonnet theory this instability cuts off large deviations from the Schwarzschild black holes. Another interesting feature of perturbations of higher curvature corrected black holes is the breakdown of the correspondence between the null geodesics and eikonal quasinormal modes \cite{Konoplya:2017wot} claimed in \cite{Cardoso:2008bp}.
We are interested to know how quasinormal modes are affected once both of the factors, the Gauss-Bonnet theory and non-linear electrodynamics are turned on.
We have chosen the form of the non-linear electrodynamics (NED) Lagrangian studied in \cite{Hyun:2019gfz} that gives automatically the Maxwellian weak-field limit, as recent results of \cite{SS1,SS2} indicate that the predictions of theories with NED Lagrangian having non-Maxwellian weak-field limit differ strongly from those for the Reissner-Nordstr\"{o}m black hole when considering the accretion disks orbiting black holes.

Another interesting problem is related to the existence of arbitrarily long lived quasinormal modes for massive fields, called quasi-resonances, which were first observed in \cite{Ohashi:2004wr} and further investigated for Schwarzschild and Kerr black holes in \cite{Konoplya:2004wg}, \cite{Konoplya:2006br}  and shown to exist not only for massive scalar, but also for massive vector \cite{Konoplya:2005hr} and Dirac fields \cite{Konoplya:2017tvu}. At the same time, there are situations in which quasi-resonances do not exist, for example, for asymptotically de Sitter black holes \cite{Konoplya:2004wg}. Even though quasi-resonances were observed in the higher dimensional Einstein-Gauss-Bonnet theories \cite{Zhidenko:2008fp}, it is not clear whether the same phenomena will take place for the case of non-linear electrodynamics.

After all, an exciting proposal was suggested by S. Hod who claimed that there is an upper bound on the damping rate of the fundamental mode in the spectrum of any black hole which is determined by the Hawking temperature \cite{Hod:2006jw}. This bound has been recently confirmed for a number of cases, including the test fields \cite{Zinhailo:2019rwd} and some channels of gravitational perturbations \cite{Cuyubamba:2016cug} in the Einstein-Gauss-Bonnet theory, but not for black holes within the non-linear electrodynamics.

Interesting model of Einstein-Gauss-Bonnet gravity in the framework of the non-linear electrodynamics has been suggested in \cite{Hyun:2019gfz}, where new five-dimensional charged (in general, asymptotically A(dS)) black hole solutions were obtained. These solutions include alongside with extremal charged black holes also regular (singularity-free) black holes. Therefore, these black holes would be an interesting model for testing a number of distinctive features of black holes via their quasinormal spectrum: checking Hod's proposal, analysis of deviation of quasinormal modes from their Eintein-Gauss-Bonnet values owing to non-linear electrodynamics, looking at the null geodesics/eikonal QNMs correspondence, etc. Having all of the above motivations in mind we analysed (both numerically and analytically) quasinormal modes of the charged  Einstein-Gauss-Bonnet black hole within non-linear electrodynamics.

The paper is organized as follows. Sec. II gives the essential information about the black hole metric under consideration. In Sec. III we deduce the wave equation for a test massive scalar field and discuss the boundary conditions. Sec. IV review briefly the WKB method we used, while in sec. V we summarize the obtained results for the quasinormal modes. Sec. VI is devoted to the checking of the Hod's proposal and sec. VII shows that there is an indication of quasi-resonances in the spectrum of massive fields. Finally, in sec. VIII we summarize the obtained results and mention some open problems.

\section{Metric and its special cases}

The five-dimensional Einstein-Gauss-Bonnet gravity coupled to a non-linear electromagnetic field is described by the action
\begin{equation}\label{action}
S=\int d^5x\sqrt{-g}\left(\frac{1}{16\pi}\left(R-2\Lambda+\alpha\mathcal{L}_{GB}\right)-\frac{1}{4\pi}\mathcal{L}\left(F\right)\right)\,,
\end{equation}
where $\Lambda$ is the negative cosmological constant related to the AdS curvature radius as
\begin{equation}\label{Lambda}
\Lambda=-\frac{6}{l^2}\,,
\end{equation}
$\mathcal{L}_{GB}$ is the Gauss-Bonnet term given by
\begin{equation}\label{GBterm}
\mathcal{L}_{GB}=R^2-4R_{\mu\nu}R^{\mu\nu}+R_{\mu\nu\rho\lambda}R^{\mu\nu\rho\lambda}\,,
\end{equation}
$\alpha$ denotes the Gauss-Bonnet coupling constant (restricted to the non-negative case due to $\alpha$ being regarded as the inverse string tension in the heterotic string theory) and $\mathcal{L}\left(F\right)$ is the function of the invariant
$$
F\equiv\frac{1}{4}F_{\mu\nu}F^{\mu\nu}\,,
$$
where
$$
F_{\mu\nu}=\partial_\mu A_\nu-\partial_\nu A_\mu
$$
is the field strength of the electromagnetic field $A_\mu$.

It turns out to be convenient to introduce a field strength
\begin{equation}\label{Pmv}
P^{\mu\nu}\equiv \frac{\partial\mathcal{L}\left(F\right)}{\partial F}F^{\mu\nu}\,,
\end{equation}
whose $\,\left(tr\right)$ component is related to the conjugate momentum of the gauge field $A_\mu$, whence the invariant $P$ has the form
\begin{equation}\label{P}
P\equiv \frac{1}{4}P_{\mu\nu}P^{\mu\nu}= \left(\frac{\partial\mathcal{L}\left(F\right)}{\partial F}\right)^2F.
\end{equation}
Using the relations
\begin{equation}\label{Hamiltonian}
\mathcal{H}\left(P\right)=2 \frac{\partial\mathcal{L}\left(F\right)}{\partial F}F-\mathcal{L}\left(F\right)\,,
\end{equation}
\begin{equation}\label{HL}
\mathcal{L}=2 \frac{\partial\mathcal{H}}{\partial P}P-\mathcal{H}\,\, , \,\, \frac{\partial\mathcal{L}}{\partial F}=\left(\frac{\partial\mathcal{H}}{\partial P}\right)^{-1}\,,
\end{equation}
and considering
\begin{equation}\label{HamiltonianNED}
\mathcal{H}\left(P\right)=P e^{-k_0\left(-P\right)^\frac{1}{3}}\,,
\end{equation}
where $k_0$ is a coupling constant of the non-linear electrodynamics, one can obtain the corresponding Lagrangian in the following expansion form:
\begin{equation}\label{Lagrangian}
\mathcal{L}\left(F\right)=F\left(1+k_0\left(-F\right)^\frac{1}{3}+\frac{23}{18}k_0^2\left(-F\right)^\frac{2}{3}+\mathcal{O}\left(k_0^3\right)\right)\,.
\end{equation}

The metric obtained in \cite{Hyun:2019gfz} reads
\begin{equation}\label{metric}
ds^2=-f(r)dt^2+\frac{dr^2}{f(r)}+r^2d\Omega_3^2\,,
\end{equation}
where
$$
f(r)=1+
$$
\begin{equation}\label{metricfunction}
+\frac{r^2}{4\alpha}\left(1-\sqrt{1-\frac{8\alpha}{l^2}+
\frac{8\alpha}{r^4}\left(m+\frac{q^2}{3k}\left( e^{-\frac{k}{r^2}}-1\right)\right)}\right)\,.
\end{equation}
The reduced mass $m$ is related to the ADM mass $M$ of the geometry by
$$
m=\frac{8}{3 \pi}M\,,
$$
the reduced charge $q$ is related to the total charge $Q$ as
$$
q=\frac{2}{\pi}Q
$$
and $k$ is related to the coupling constant $k_0$ of the non-linear electrodynamics by
$$
k=k_0\left(\frac{q^2}{2}\right)^{\frac{1}{3}}\,.
$$

In the limit $l\rightarrow \infty$ we have asymptotically flat space-time, so that the metric function has the form
\begin{equation}\label{metricfunctionGB}
f(r)=1+\frac{r^2}{4\alpha}\left(1-\sqrt{1+
\frac{8\alpha}{r^4}\left(m+\frac{q^2}{3k}\left( e^{-\frac{k}{r^2}}-1\right)\right)}\right)\,.
\end{equation}

If we take additional limit $\alpha \rightarrow 0$ the metric function reads
\begin{equation}\label{metricfunctionK}
f(r)=1-\frac{m}{r^2}-\frac{q^2}{3kr^2}\left( e^{-\frac{k}{r^2}}-1\right)\,.
\end{equation}

Note that if $k \rightarrow 0$, usual Maxwell electrodynamics is restored and from (\ref{metricfunctionGB}) we obtain the Wiltshire black hole solution \cite{Wiltshire:1988uq}
\begin{equation}\label{metricfunctionW}
f(r)=1+\frac{r^2}{4\alpha}\left(1-\sqrt{1+
\frac{8\alpha m}{r^4}-\frac{8\alpha q^2}{3r^6}}\right)\,
\end{equation}
and from (\ref{metricfunctionK}) we obtain the Reissner-Nordstr\"{o}m black hole
\begin{equation}\label{metricfunctionRN}
f(r)=1-\frac{m}{r^2}+\frac{q^2}{3r^4}\,,
\end{equation}
with $m=2M$ and $q=Q\sqrt{3}$.

Finally, in the limit $q \rightarrow 0$ from (\ref{metricfunctionW}) we get the Boulware-Deser black hole solution \cite{Boulware:1985wk}
\begin{equation}\label{metricfunctionBD}
f(r)=1+\frac{r^2}{4\alpha}\left(1-\sqrt{1+
\frac{8\alpha m}{r^4}}\right)\,
\end{equation}
and from (\ref{metricfunctionRN}) we get the Schwarzschild solution
\begin{equation}\label{metricfunctionS}
f(r)=1-\frac{m}{r^2}\,.
\end{equation}

Our analysis of the Gauss-Bonnet gravity combined with the "exponential" non-linear electrodynamics is restricted to the asymptotically flat spacetime case with $\Lambda=0$. Our results are thus relevant for the metric given by Eq. (\ref{metricfunctionGB}) and all its subclasses.

\section{The wave equation}

We consider scalar perturbations of a black hole space-time. For the scalar test field $\Phi$ of mass $\mu$ these perturbations can be represented by general Klein-Gordon equation
\begin{equation}\label{KG}
\left(\Box - \mu^2\right)\Phi = 0\,.
\end{equation}
This equation can be written explicitly in the background of the metric $g_{\mu \nu}$:
\begin{equation}\label{KGg}
\frac{1}{\sqrt{-g}}\partial_\nu \left(g^{\mu \nu}\sqrt{-g}\partial_\mu\Phi\right)-\mu^2\Phi=0\,.
\end{equation}
Taking the function $\Phi$ in the form
$$
\Phi\left(t,\,r,\,\theta,\,\phi\right)=e^{\pm i \omega t}Y_\ell\left(\theta,\,\phi\right)\Psi\left(r\right)/r \,,
$$
where $\ell$ is the multipole quantum number for the usual spherical harmonics $Y_\ell\left(\theta,\,\phi\right)$ \footnote{We do not write down the full spherical harmonics with both indices $Y_\ell^m$, because for our spherically symmetric case there is a parameter $m$ degeneracy.}, we obtain Schr\"{o}dinger-like equation
\begin{equation}\label{wave-equation}
\frac{d^2\Psi}{dr_*^2}+\left(\omega^2-V(r)\right)\Psi=0,
\end{equation}
with respect to the "tortoise coordinate" $r_*$, mapping the event horizon to $-\infty$,
\begin{equation}
dr_*=\frac{dr}{f(r)}.
\end{equation}
Solving this equation with strictly fixed boundary conditions: only incoming waves at the horizon ($r_*\rightarrow -\infty$) and only outgoing waves at infinity ($r_*\rightarrow +\infty$), we find quasinormal modes of the black hole. That means we obtain a discrete set of complex values for the frequencies $\omega$, where the real part is the oscillation frequency and imaginary part is proportional to the damping rate of the oscillations.

\begin{figure}
\resizebox{\linewidth}{!}{\includegraphics*{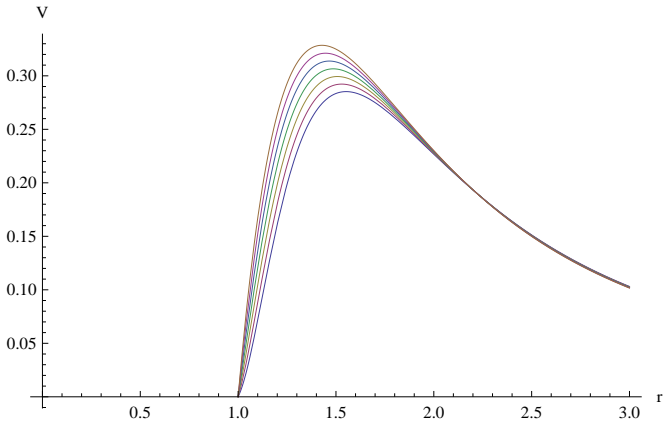}}
\caption{Scalar potential for $l=0$, $q=1.5$, from $k=0$ (down) to $k=0.6$ (up).}\label{fig:potL0}
\end{figure}

The five-dimensional scalar effective potential for the metric (\ref{metricfunctionGB}) is given by (see for example, \cite{Konoplya:2004xx})
\begin{equation}\label{potential}
V(r)=f(r)\left(\frac{\ell\left(\ell+ 2\right)}{r^2}+\frac{3}{4r^2}f(r)+\frac{3}{2r}f'(r)+\mu^2\right)\,
\end{equation}
and has the form of the potential barrier (see Fig. \ref{fig:potL0}).

\section{The methods}

To study the quasinormal modes we use two methods.

\subsection{WKB method}

The WKB method, which was first used for finding quasinormal modes in the work of Schutz and Will \cite{Schutz:1985zz} and at the first order reproduced the earlier result of Mashhoon \cite{Mashoon}, proved to be very effective and gained rapid development in numerous papers. It quickly became extremely popular due to its good accuracy and automaticity.

For finding quasinormal frequencies we use higher-order WKB formula \cite{Mashoon,Schutz:1985zz,Iyer:1986np,Konoplya:2003ii,Matyjasek:2017psv,Konoplya:2019hlu}:
\begin{eqnarray}\label{WKBformula-spherical}
\omega^2&=&V_0+A_2(\K^2)+A_4(\K^2)+A_6(\K^2)+\ldots\\\nonumber&-&\imo \K\sqrt{-2V_2}\left(1+A_3(\K^2)+A_5(\K^2)+A_7(\K^2)\ldots\right),
\end{eqnarray}
where $\K$ takes half-integer values
\begin{eqnarray}\label{QNMsK}
\K &=& \left\{
\begin{array}{ll}
 +n+\frac{1}{2}, & \re{\omega}>0; \\
 -n-\frac{1}{2}, & \re{\omega}<0; \phantom{\frac{{}^{Whitespace}}{}}
\end{array}
\right.\\\nonumber
&&\qquad\quad\qquad n=0,1,2,3\ldots.
\end{eqnarray}
The corrections $A_k(\K^2)$ of order $k$ to the eikonal formula are polynomials of $\K^2$ with rational coefficients and depend on the values $V_2, V_3\ldots$ of higher derivatives of the potential $V(r)$ in its maximum (but not on the value $V_0$ of the potential $V(r)$ itself), whence the righthand-side of (\ref{WKBformula-spherical}) does not depend on the value of $\omega$.

It is well known that, as WKB method converges only asymptotically, the higher order of the WKB formula does not guarantee improving of the results. However, more on reaching the asymptotic WKB regime can be seen in \cite{Hatsuda:2019eoj}.  To increase accuracy of the higher-order WKB formula, we follow Matyjasek and Opala \cite{Matyjasek:2017psv} and use Padé approximants \cite{PadeApproximation}. For the order $k$ of the WKB formula (\ref{WKBformula-spherical}) we define a polynomial $P_k(\epsilon)$ as follows
\begin{eqnarray}\nonumber
  P_k(\epsilon)&=&V_0+A_2(\K^2)\epsilon^2+A_4(\K^2)\epsilon^4+A_6(\K^2)\epsilon^6+\ldots\\&-&\imo \K\sqrt{-2V_2}\left(\epsilon+A_3(\K^2)\epsilon^3+A_5(\K^2)\epsilon^5\ldots\right),\label{WKBpoly}
\end{eqnarray}
whence the squared frequency is obtained for $\epsilon=1$:
$$\omega^2=P_k(1).$$

For the polynomial $P_k(\epsilon)$ we construct rational functions
\begin{equation}\label{WKBPade}
P_{\tilde{n}/\tilde{m}}(\epsilon)=\frac{Q_0+Q_1\epsilon+\ldots+Q_{\tilde{n}}\epsilon^{\tilde{n}}}{R_0+R_1\epsilon+\ldots+R_{\tilde{m}}\epsilon^{\tilde{m}}},
\end{equation}
called Padé approximants, with $\tilde{n}+\tilde{m}=k$, such that near $\epsilon=0$
$$P_{\tilde{n}/\tilde{m}}(\epsilon)-P_k(\epsilon)={\cal O}\left(\epsilon^{k+1}\right).$$

It appears in practice that for finding fundamental mode ($n=0$) Padé approximants with $\tilde{n}\approx\tilde{m}$ give the best approximation. In \cite{Matyjasek:2017psv}, $P_{6/6}(1)$ and $P_{6/7}(1)$ were compared to the 6th-order WKB formula $P_{6/0}(1)$. In \cite{Konoplya:2019hlu} it has been observed that usually even $P_{3/3}(1)$, i.~e. a Padé approximation of the 6th-order, gives a more accurate value for the squared frequency than $P_{6/0}(1)$.
We use this observation to find appropriate Padé partition for different $\ell$ and extrapolate it to our case. Although the WKB method is known to work properly for $\ell> n$ \cite{Iyer:1986np}, using of a suitable Padé approximation often gives rather good results even for $\ell=0$.

\subsection{Time domain method}

In the cases when the WKB method cannot be applied (e. g. if the effective potential has more than two turning points) or when it gives the error which is larger than effect (as sometimes is for $\ell=n=0$), we have to use the time domain integration of the perturbation equations instead. We shall integrate the wave-like equation rewritten in terms of the light-cone variables $u=t-r_*$ and $v=t+r_*$. The appropriate discretization scheme was suggested in \cite{Gundlach:1993tp}:
$$
\Psi\left(N\right)=\Psi\left(W\right)+\Psi\left(E\right)-\Psi\left(S\right)-
$$
\begin{equation}\label{Discretization}
-\Delta^2\frac{V\left(W\right)\Psi\left(W\right)+V\left(E\right)\Psi\left(E\right)}{8}+{\cal O}\left(\Delta^4\right)\,,
\end{equation}
where we used the following designations for the points:
$N=\left(u+\Delta,v+\Delta\right)$, $W=\left(u+\Delta,v\right)$, $E=\left(u,v+\Delta\right)$ and $S=\left(u,v\right)$. The initial data are given on the null surfaces $u=u_0$ and $v=v_0$. As the $\ell=0$ mode is characterized by a relatively short period of quasinormal ringing, it is difficult to extract the values of the frequencies with good accuracy, so that we write down only three digits after the point for $\omega$.

\begin{figure*}
\resizebox{\linewidth}{!}{\includegraphics*{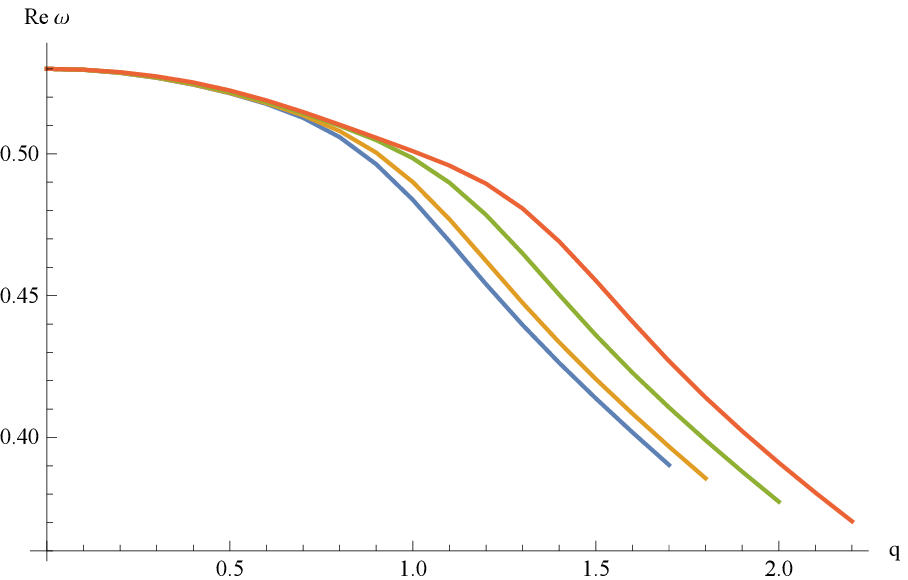}\includegraphics*{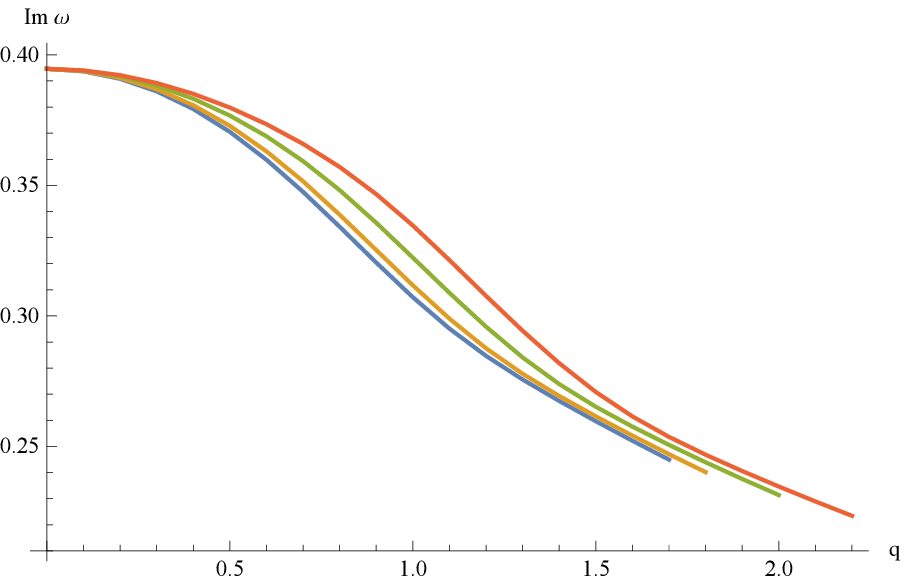}}
\caption{Real part (left panel) and imaginary part (right panel) of the fundamental ($n=0$) quasinormal mode depending on $q$, for $l=0$, $k=0, 0.1, 0.3, 0.5$ (lines are higher with higher $k$), calculated with the WKB method.}\label{fig:L0Re}
\end{figure*}

\begin{figure*}
\resizebox{\linewidth}{!}{\includegraphics*{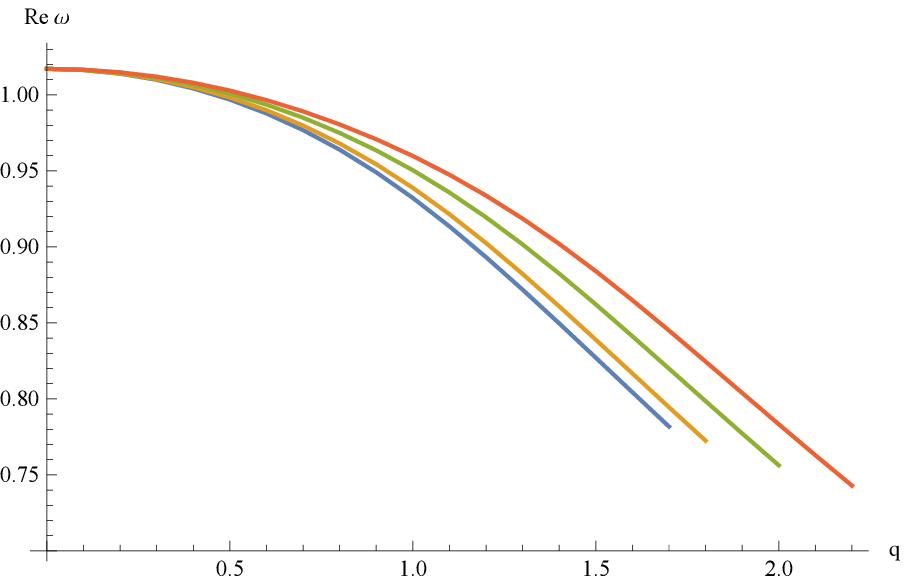}\includegraphics*{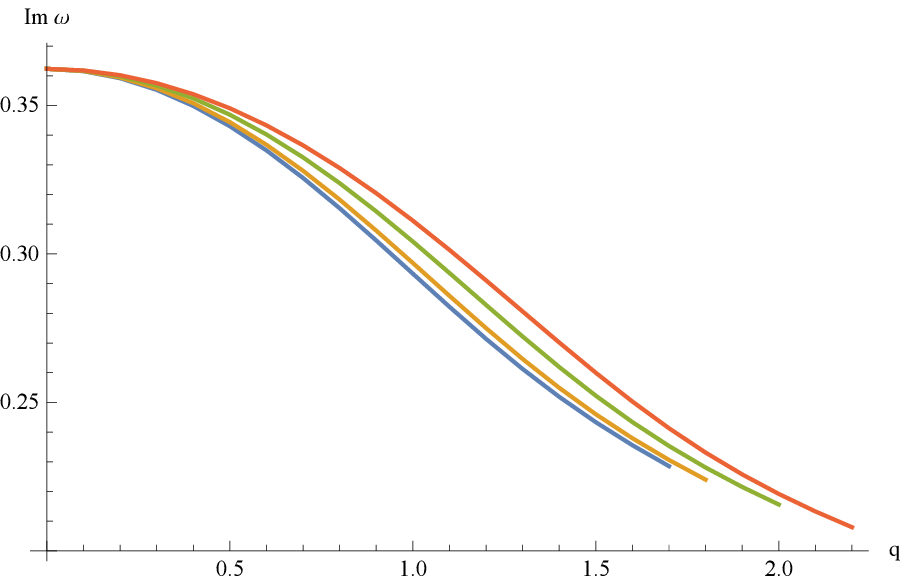}}
\caption{Real part (left panel) and imaginary part (right panel) of the fundamental ($n=0$) quasinormal mode depending on $q$, for $l=1$, $k=0, 0.1, 0.3, 0.5$ (lines are higher with higher $k$), calculated with the WKB method.}\label{fig:L1Re}
\end{figure*}

\begin{figure*}
\resizebox{\linewidth}{!}{\includegraphics*{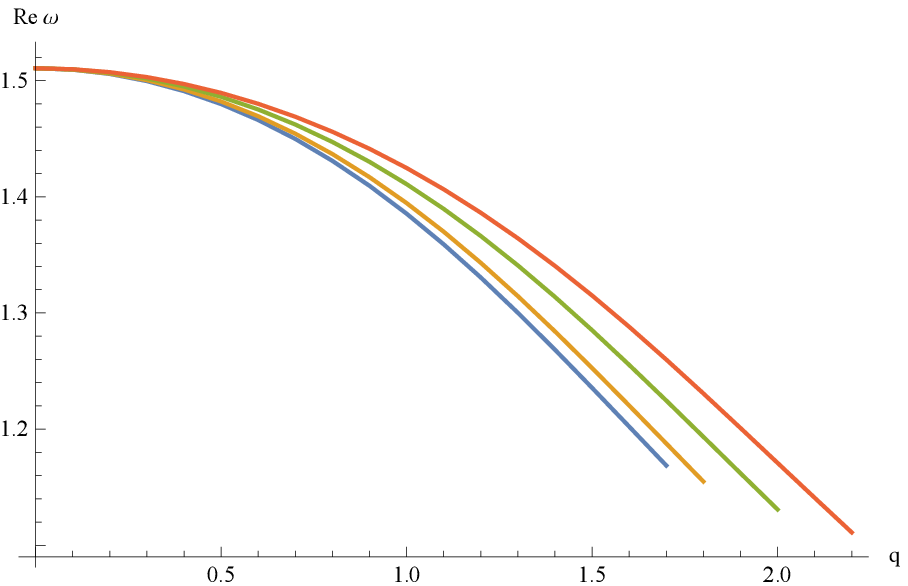}\includegraphics*{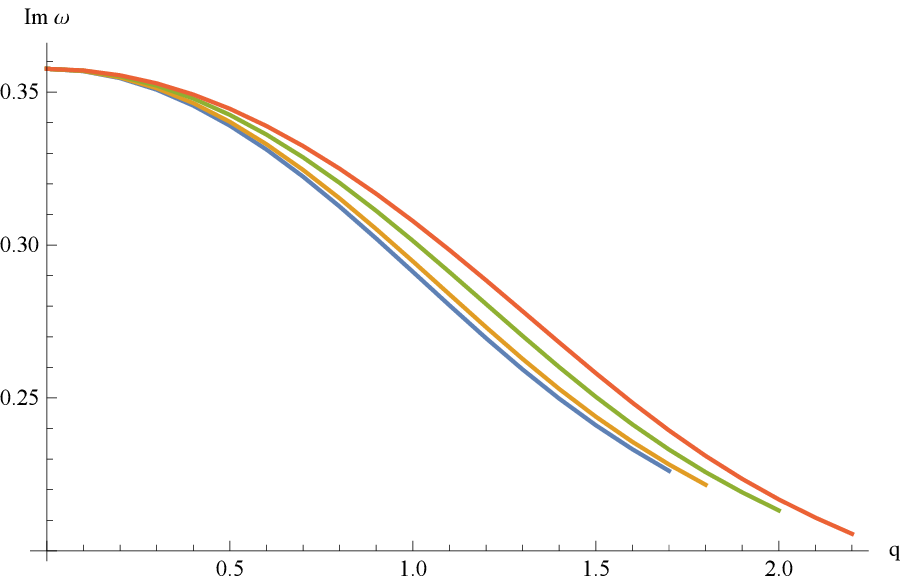}}
\caption{Real part (left panel) and imaginary part (right panel) of the fundamental ($n=0$) quasinormal mode depending on $q$, for $l=2$, $k=0, 0.1, 0.3, 0.5$ (lines are higher with higher $k$), calculated with the WKB method.}\label{fig:L2Re}
\end{figure*}

\begin{figure*}
\resizebox{\linewidth}{!}{\includegraphics*{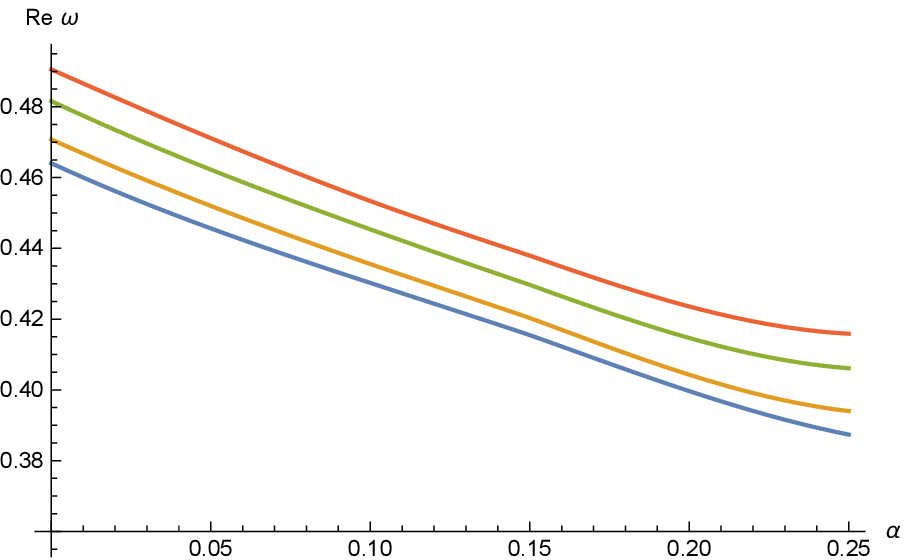}\includegraphics*{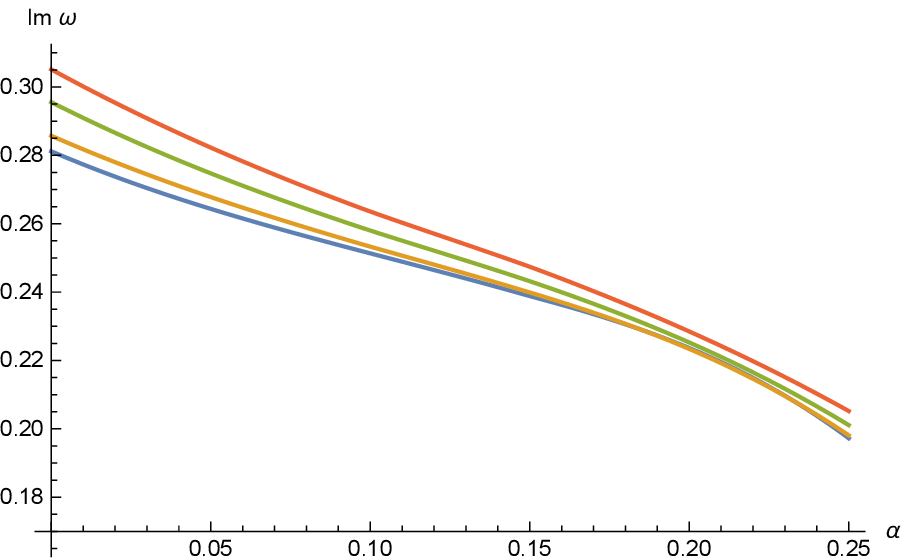}}
\caption{Real part (left panel) and imaginary part (right panel) of the fundamental ($n=0$) quasinormal mode depending on $\alpha$, for $l=0$, $q=1.2$, $k=0, 0.1, 0.3, 0.5$ (lines are higher with higher $k$), calculated with the time-domain method.}\label{fig:AlphaL0ReTD}
\end{figure*}

\begin{figure*}
\resizebox{\linewidth}{!}{\includegraphics*{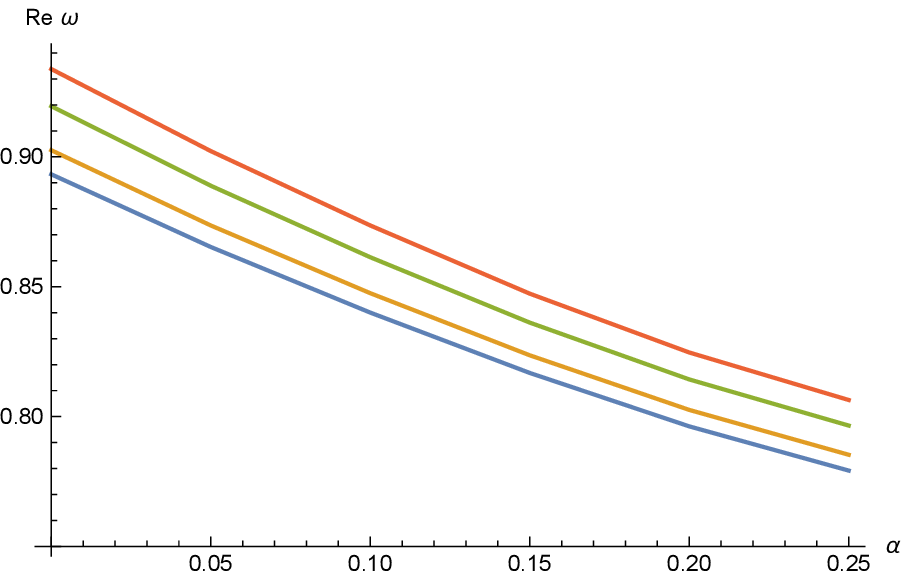}\includegraphics*{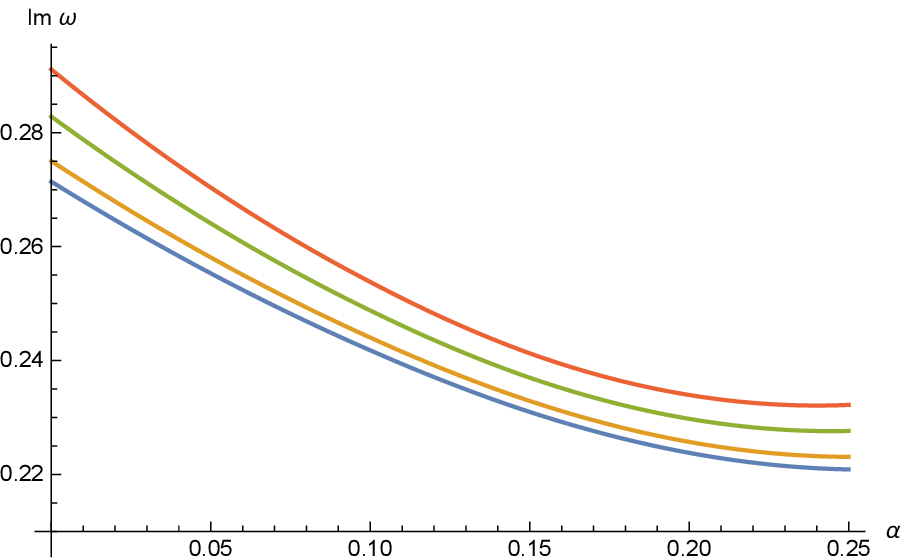}}
\caption{Real part (left panel) and imaginary part (right panel) of the fundamental ($n=0$) quasinormal mode depending on $\alpha$, for $l=1$, $q=1.2$, $k=0, 0.1, 0.3, 0.5$ (lines are higher with higher $k$), calculated with the WKB method.}\label{fig:AlphaL1Re}
\end{figure*}

\begin{figure*}
\resizebox{\linewidth}{!}{\includegraphics*{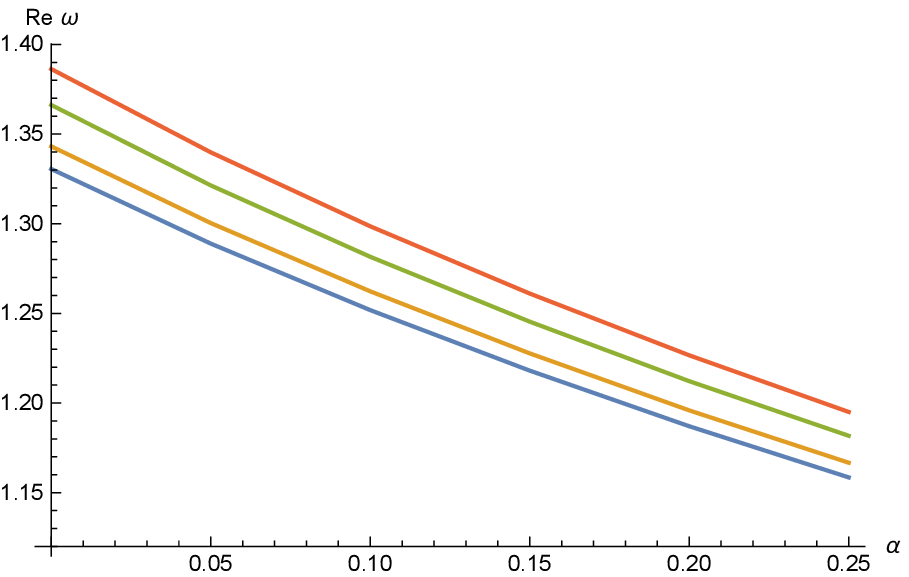}\includegraphics*{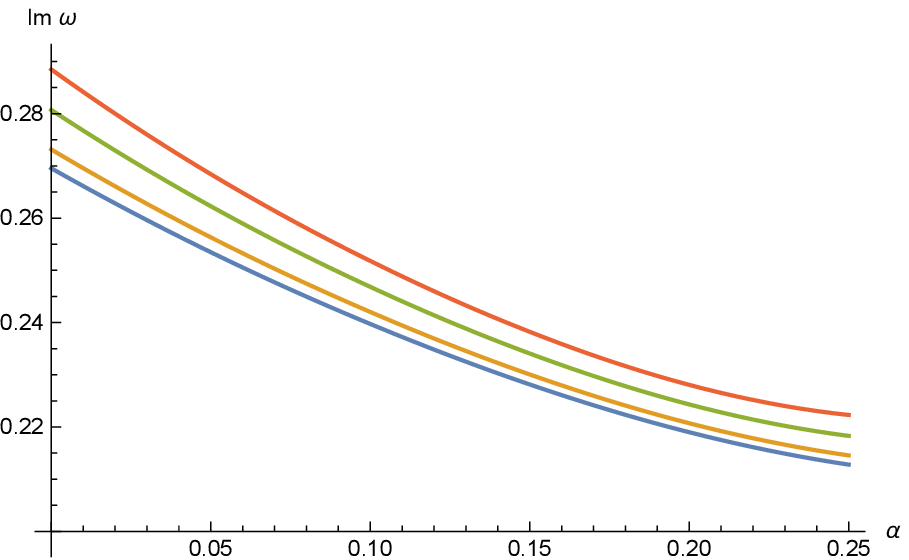}}
\caption{Real part (left panel) and imaginary part (right panel) of the fundamental ($n=0$) quasinormal mode depending on $\alpha$, for $l=2$, $q=1.2$, $k=0, 0.1, 0.3, 0.5$ (lines are higher with higher $k$), calculated with the WKB method.}\label{fig:AlphaL2Re}
\end{figure*}

\begin{center}
\begin{table*}
\begin{tabular}{p{2cm}cc}
\multicolumn{3}{c}{{\bf Table 1.} Quasinormal modes obtained by WKB and time-domain methods.} \\
\hline
k & WKB & Time domain  \\[3pt]  \hline \\[-5pt]
\multicolumn{3}{c}{$\ell=0$, $q=1.2$} \\
\hline \\[-5pt]
0 & $0.454055-0.284700i$ & $0.464 -
0.282i$  \\[5pt]
0.1 & $0.462254-0.287604i$ & $0.471 -
0.286i$  \\[5pt]
0.3 & $0.478495-0.295874i$ & $0.482 -
0.296i$ \\[5pt]
0.5 & $0.489508-0.307674i$ & $0.490 - 0.306i$ \\[5pt]
\hline
\multicolumn{3}{c}{$\ell=1$, $\alpha=0.1$, $q=1.2$} \\
\hline
0 & $0.840041-0.241797i$ & $0.839 -
0.242i$  \\[5pt]
0.1 & $0.847513-0.244032i$ & $0.846 -
0.244i$  \\[5pt]
0.3 & $0.861360-0.248789i$ & $0.860 -
0.249i$ \\[5pt]
0.5 & $0.873591-0.253796i$ & $0.872 - 0.254i$ \\[5pt]
\hline
\multicolumn{3}{c}{$\ell=2$, $\alpha=0.1$, $q=1.2$} \\
\hline
0 & $1.251824-0.239738i$ & $1.251 -
0.240i$  \\[5pt]
0.1 & $1.262301-0.242034i$ & $1.262 -
0.242i$  \\[5pt]
0.3 & $1.281566-0.246844i$ & $1.281 -
0.247i$ \\[5pt]
0.5 & $1.298571-0.251814i$ & $1.298 - 0.252i$ \\[5pt]
\hline
\end{tabular}
\end{table*}
\end{center}

\section{Quasinormal modes}

Both WKB method and time-domain integration have their own advantages and drawbacks. The time-domain integration always gives an accurate time-domain profile but the problem is to extract correctly the value of the quasinormal mode, so that we trust only two or three digits after the point. On the other hand, the WKB method is known to converge only asymptotically and its accuracy can be told only by comparing the values obtained for different orders. Nevertheless, as can be seen from Table 1, in our case the WKB data and results of time-domain integration are in a good concordance in the common region of their applicability.

We start with the limit $\alpha \rightarrow 0$ and find quasinormal modes depending on $q$ (starting from zero) for different $k$ (starting from zero). That means the lowest line on Figs. \ref{fig:L0Re} -- \ref{fig:L2Re} (corresponding to $k=0$) shows QNMs for the Schwarzschild black hole ($k=0$, $q=0$) and for the Reissner-Nordstr\"{o}m black hole ($k=0$, $q\neq 0$), the higher lines standing for the nonlinear electrodynamics of the charged (excluding the values on the vertical axis) black holes.

Adding Gauss-Bonnet term, we fix $q=1.2$ and take $\alpha$ not greater than $2.5$ to avoid the eikonal instability, which is known for the five-dimensional GB theory for neutral black hole \cite{Konoplya:2008ix,Dotti:2005sq,Gleiser:2005ra} and for charged case within Maxwell electrodynamics
\cite{Takahashi:2012np,Takahashi:2011qda} and is, therefore, highly expected for large values of the GB coupling constant in our case. Thus the lowest line on Figs. \ref{fig:AlphaL0ReTD} -- \ref{fig:AlphaL2Re} (corresponding to $k=0$) shows QNMs for the Reissner-Nordstr\"{o}m black hole ($k=0$, $\alpha=0$) and for the Wiltshire black hole ($k=0$, $\alpha\neq 0$). Note that in this case (with the Gauss-Bonnet term added) for $\ell=0$ we used time-domain method as it gives more accurate values for quasinormal frequencies.

The obtained results show that when the Gauss-Bonnet coupling is turned on, the real oscillations frequency and the damping rate decrease, while the coupling due to non-linear electrodynamics increases real and imaginary parts of the quasinormal modes. This influence is rather noticeable and may come to tens of percents.

In the limit $\alpha \rightarrow 0$, that is in the case of the metric (\ref{metricfunctionK}), using the general approach suggested in \cite{Churilova:2019jqx}, we obtain the following eikonal formula for the squared frequency:
\begin{widetext}
$$ \omega^2 = \frac{-3 k \ell (\ell+2) q^2+6 \ell^2 r_0^2 \left(-3 m r_0^2+q^2+3 r_0^4\right)+12 L r_0^2 \left(-3 m r^2+q^2+3 r_0^4\right)}{18 r_0^8} + \frac{i (2 n+1)}{6 \sqrt{3}} \frac{\ell (\ell+2)}{r_0^{22}} \sqrt{P},$$
where
 $$  P= 30 k^3 q^6+k^2 q^4 r_0^2 \left(63 \left(6 m r_0^2-5 r_0^4\right)-151 q^2\right)+4 k q^2 r_0^4 \left(9 \left(40 m^2 r_0^4-63 m r_0^6+24 r_0^8\right)+ q^2 \left(246 r_0^4-303 m r_0^2\right)+62
   q^4\right)-$$
\begin{equation}
12 r_0^6 \left(3 q^2 \left(59 m^2 r_0^4-88 m r_0^6+31 r_0^8\right)+9 r_0^6 \left(-14 m^3+29 m^2 r_0^2-18 m r_0^4+3 r_0^6\right)+q^4 \left(61 r_0^4-78 m r_0^2\right)+11 q^6\right).
\end{equation}
\end{widetext}
where $k$ is sufficiently small and $\ell$ is large. Here $r_0$ is the position of the maximum of the effective potential. Our numerical computations both in time and frequency domains confirm the validity of the above eikonal formula when the multipole number is sufficiently large.

\begin{figure*}
\resizebox{\linewidth}{!}{\includegraphics*{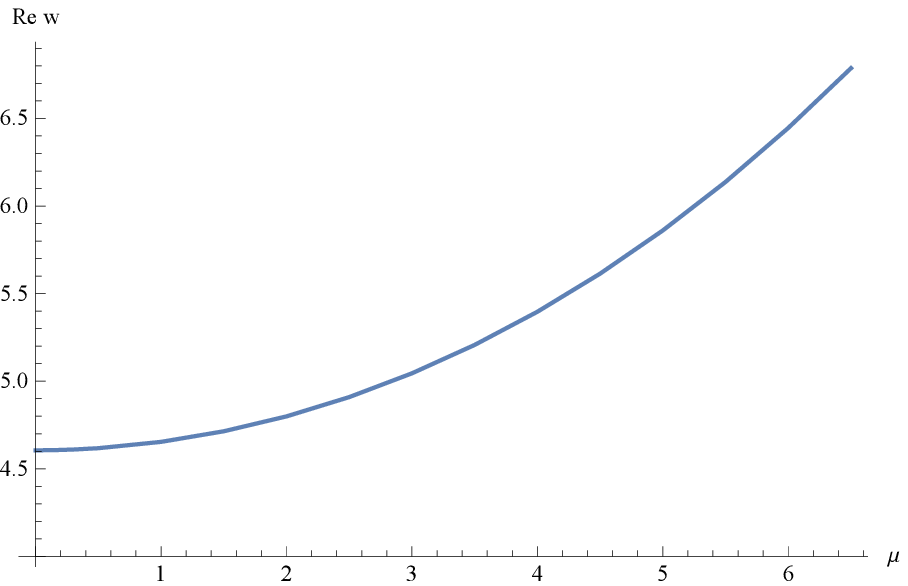}\includegraphics*{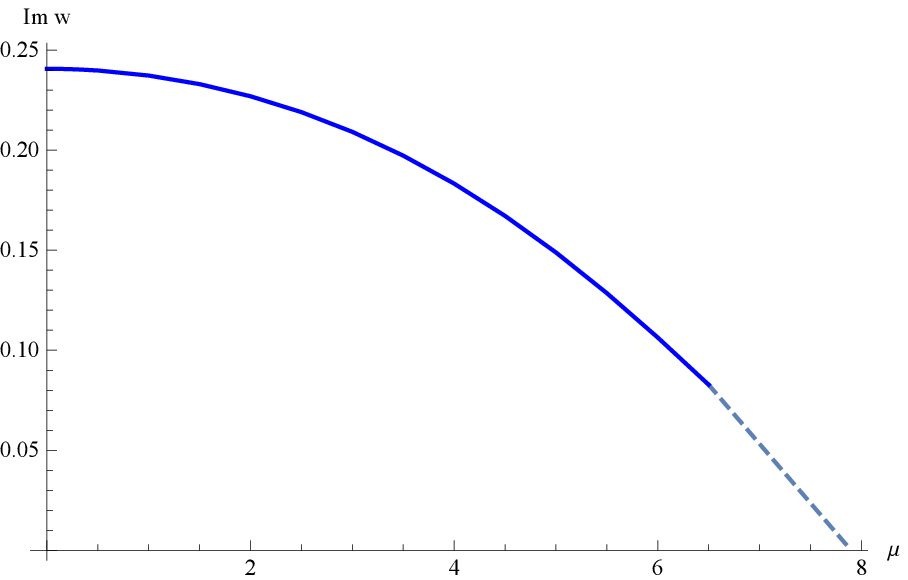}}
\caption{Real part (left panel) and imaginary part (right panel) of the fundamental ($n=0$) quasinormal mode depending on $\mu$, for $l=10$, $q=1.2$, $k=0.1$, $\alpha=0.1$, calculated with the WKB method.}\label{fig:MiuRe}
\end{figure*}

\begin{figure}
\resizebox{\linewidth}{!}{\includegraphics*{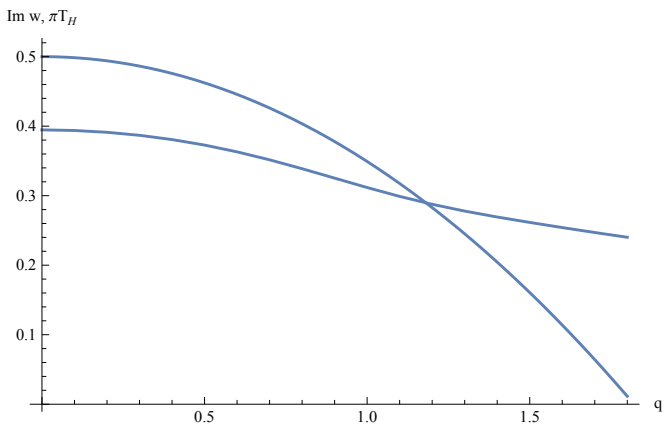}}
\caption{Imaginary part of the fundamental ($n=0$) quasinormal mode (the line which starts lower), calculated with the WKB method, and Hawkimg temperature depending on $q$, for $l=0$, $k=0.1$.}\label{fig:TH}
\end{figure}

\section{The Hod's proposal}

In \cite{Hod:2006jw} Hod set forth the following statement about the damping rate of the fundamental oscillation: in the spectrum of any black hole's quasinormal modes there always must exist a frequency $\omega$ such that
\begin{equation}\label{Hod}
\left|Im\left(\omega\right) \right|\leq \pi T_H\,,
\end{equation}
where $T_H$ is the Hawking temperature of the black hole. In this paper \cite{Hod:2006jw} Hod considered not only four-dimensional black holes but also high-dimensional black holes and even asymptotically AdS black holes. In fact his claim (\ref{Hod}) is general and concerns all the black holes.

The Hawking temperature for our case can be written as
\begin{equation}\label{TH}
T_H=\left.\frac{1}{4 \pi}f'\left(r\right)\right|_{r=r_h}\,,
\end{equation}
where $r_h$ is the radius of the event horizon.

Although, as one can see on Fig. \ref{fig:TH}, starting from certain value of parameter $q$, the Hod's conjecture seems to be broken, it is not the final answer because here we consider scalar test fields only. However, the fact that the test scalar field formally violates the inequality motivates further study of gravitational spectrum of these black holes.

\section{Quasiresonance}

It was shown in \cite{Konoplya:2004wg} that for the massive scalar field there exists a phenomenon of so-called quasiresonance. This means that when the mass of the field increases the damping rate of the lower overtones decreases, which causes the appearance of the infinitely long lived modes, and after some threshold value of the mass the lower overtones disappear from the spectrum. On the contrary, all the remaining higher overtones are still damping. For the first time decreasing of the damping rate (though for different boundary conditions corresponding to bound states) when the mass is increased was presented in \cite{Starobinsky:1973aij}.

Fig. \ref{fig:MiuRe}, showing dependance of the real and imaginary parts of the fundamental mode on the mass $\mu$, indicates that for the considered case of the massive scalar field in the five-dimensional Einstein-Gauss-Bonnet gravity coupled to a nonlinear electrodynamics the phenomenon of quasiresonance exists.

The WKB method we used here cannot be used in the regime of quasi-resonances, but, as was shown in \cite{Konoplya:2017tvu}, it still works when $\ell$ is much larger than $\mu M$, so that the extrapolation of the WKB data can clearly indicate the existence of quasi-resonances.

\section{Conclusions}

We found the quasinormal modes of a massive scalar field in the background of the five-dimensional, charged Einstein-Gauss-Bonnet black hole coupled to a nonlinear electrodynamics and answered the number of appealing questions:

\begin{itemize}
\item We computed quasinormal modes of a test massive scalar field in the background of the charged asymptotically flat Einstein-Gauss-Bonnet black hole within the non-linear electrodynamics. We used the higher order WKB method with Pad\'{e} approximants and the time-domain integration. Both methods are in good concordance in the range of their validity. Thus we successfully tested the method of Pad\'{e} approximation, suggested in \cite{Matyjasek:2017psv}. In the eikonal regime an analytical treatment was done for QNMs.
\item We found that the inequality proposed by Hod is formally violated for the test scalar field in this theory, which means that the gravitational perturbations must be considered to check the proposal.
\item It has been shown that there is a clear indication of existence of arbitrarily long lived quasinormal modes, quasi-resonances, which means that, apparently, the scope of phenomenon of quasi-resonances is broader than it was known before.
\item When the Gauss-Bonnet coupling is turned on, the real oscillations frequency and the damping rate decrease, while the coupling due to non-linear electrodynamics increases real and imaginary parts of the quasinormal modes. This influence is rather noticeable and may come to tens of percents.
\end{itemize}

\acknowledgments{The authors acknowledge  the  support  of  the  grant  19-03950S of Czech Science Foundation ($GA\check{C}R$). M. S. C. acknowledges hospitality and support of Silesian University in Opava and Roman Konoplya for useful discussions.}

\end{document}